\documentclass[pra,twocolumn,showpacs,
	floatfix, reprint]{revtex4-2}
	\usepackage[usenames,dvipsnames]{color}
	\usepackage{graphicx}
	\usepackage{color}
	\usepackage{booktabs}
	\usepackage{amsmath}
	\usepackage{multirow}
	\usepackage{amssymb,amsmath,verbatim,ulem}
	\usepackage{tikz}

	\newcommand{\ket}[1]{|{#1}\rangle}
	\newcommand{\bra}[1]{\langle{#1}|}

	\definecolor{lightgrey}{RGB}{150,150,150} 

	\usepackage{hyperref}
	\hypersetup{
	  colorlinks   = true, 
	  urlcolor     = blue, 
	  linkcolor    = blue, 
	  citecolor   = blue 
	}
	
\begin{document}
	
	\title{Role of spontaneously generated coherence (SGC) in laser cooling of atoms}

	
	\author{Rajnandan Choudhury Das}
	 \author{Samir Khan}
	 \author{Thilagaraj R}
	\author{Kanhaiya Pandey}%
	\email{kanhaiyapandey@iitg.ac.in}
	\affiliation{%
	 Department of Physics, Indian Institute of Technology Guwahati, Guwahati, Assam 781039, India
	}%
	
	\date{\today}
	
	\begin{abstract} 
The well-known sub-Doppler polarization gradient cooling in type-I transition ($F_e=F_g+1$) is caused by red-detuned lasers. On the other hand, in type-II transition ($F_e\le F_g$), sub-Doppler cooling takes place through blue-detuned lasers. This opposite behavior for the two types of transitions is due to SGC. In the absence of SGC, both types of transitions show blue-detuned cooling. In this work, we experimentally and theoretically demonstrate blue-detuned cooling for both types of transitions in $^{\textrm{87}}$Rb. For completeness, we compare the temperatures in various configurations.
\end{abstract}
	
	\maketitle

\section{Introduction}
Spontaneouly generated coherence (SGC) plays an important role in spectroscopy and has been extensively studied in multilevel systems \cite{Menon1998PRA,Lezama1999PRA,Taichenachev1999PRA,Xu2003OptComm,Paspalakis1998OptComm,Zheng2011PRA,Zheng2016JPhyB,Silatan2023ScRep}. Since the behaviour of laser cooling has a direct connection with the atomic spectrum profile, SGC also plays a crucial role.    

In alkali atoms, laser cooling is generally realized using type-I transition, i.e., $F_g$ $\rightarrow$ $F_e=F_g+1$ (where $F_g$ and $F_e$ are ground and excited state angular momentum). In type-I transition, both Doppler and sub-Doppler cooling require red-detuned lasers. In contrast, for type-II transition ($F_g$ $\ge$ $F_e$), Doppler cooling requires red-detuned lasers, while sub-Doppler cooling requires blue-detuned lasers \cite{Devlin2016NewJP}. The opposite behavior of these two types of transitions is attributed to SGC. Red-detuned sub-Doppler cooling is due to SGC which is prominent if the magnetic sub-levels are degenerate and is therefore fragile to the magnetic field \cite{McGilligan2017SciRep,Chin2017PRA, Walhout1996PRA, Walhout1992JosaB, Devlin2016NewJP, Tarbutt2018PRL}. 
Achieving low temperatures involves carefully nullifying the magnetic field, a task not feasible in the MOT phase.
It is known that blue-detuned cooling occurs in MOT at type-II transition and has been demonstrated in Rb with significant advantages \cite{Tarbutt2018PRL,Tarbutt2018PRA,Piest2022Rb}. Blue-detuned MOTs have also been demonstrated in molecules and are crucial for cold molecule experiments \cite{Burau2023PRL,Xu2023PRA,jorapur2023high,li2023bluedetuned}. In this work, we show that blue-detuned cooling in MOT is possible even in type-I transition.

In the experiment, we utilize a narrow open transition in $^{87}$Rb for MOT \cite{rajnandan2023, DingNIST2022,rajnandan2024continuous}, similar to Li \cite{Hulet2011, Dieckmann2014}, K \cite{Thywissen2011}, Ca \cite{Hollberg2001, Hollberg2003}, Sr \cite{Makoto1999, Wilk2014EPJD}, Yb \cite{Yabuzaki1999, Natarajan2010}, Dy \cite{Benjamin2011, Pfau2014}, Er \cite{Jabez2008, Ferlaino2012, Boong2020}, Cd \cite{Katori2019}, Eu \cite{Mikio2021}, etc. The MOT for $^{87}$Rb atoms is realized at a narrow open transition at 420 nm, which has a 4-5 times lower theoretical Doppler temperature (33~$\mu$K) than the regularly used broad transition at 780 nm. Since the blue transition is open, we observe that the repumper laser plays a crucial role in lowering the temperature of the narrow-line MOT. We observed that the blue-detuned repumper causes further cooling in both type-I ($F_g=1$ $\rightarrow$ $F_e=2$) and type-II ($F_g=1$ $\rightarrow$ $F_e=1$) transitions in the MOT. We have also explored various other configurations that produce both red-detuned and blue-detuned blue MOTs of Rb.

In this work, we undertake a comprehensive numerical investigation employing a density matrix analysis to explore the impact of SGC on the sub-Doppler force within an atomic system characterized by the $F_g = 1 \rightarrow F_e = 2$ transition. Our primary objective is to assess the feasibility of blue detuned laser cooling in MOT. Subsequently, through experimental exploration, we investigate various configurations of blue-detuned MOTs for Rb at a narrow transition on the D$_1$ and D$_2$ lines. Our results, augmented by the density matrix analysis, demonstrate the efficacy of blue-detuned laser cooling even in type-I MOTs, achieving sub-Doppler temperatures as low as $24~\mu$K in the D$_1$ MOT and $31~\mu$K in the D$_2$ MOT.

\section{Theory}
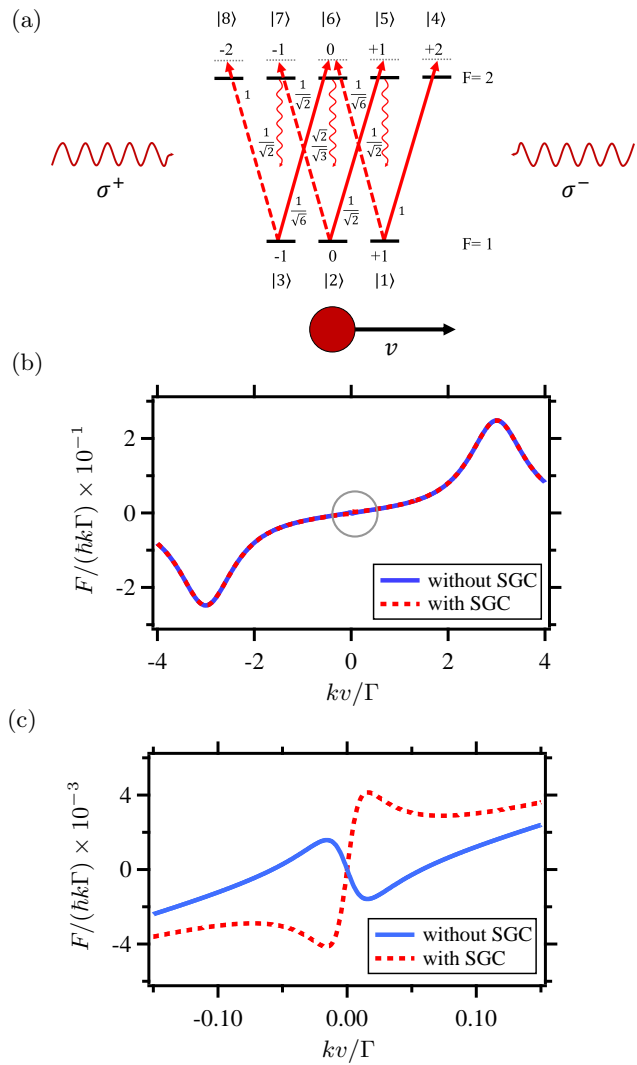
\begin{figure}
	\centering
	\includegraphics[width=1\linewidth]{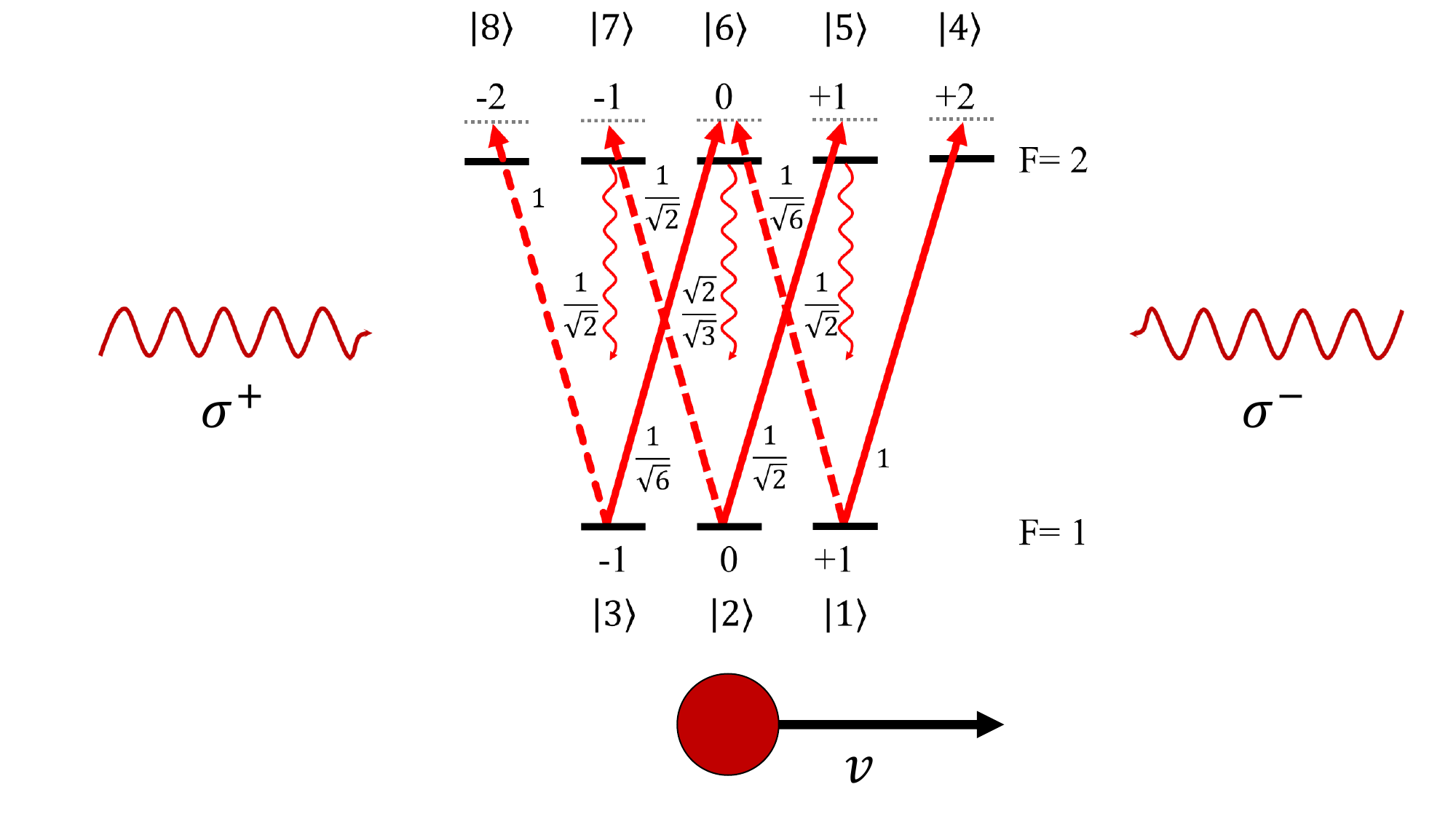}
	\begin{picture}(0,0)
		\put(-20,255){(a)}
	\end{picture}
	  \includegraphics[width=0.8\linewidth]{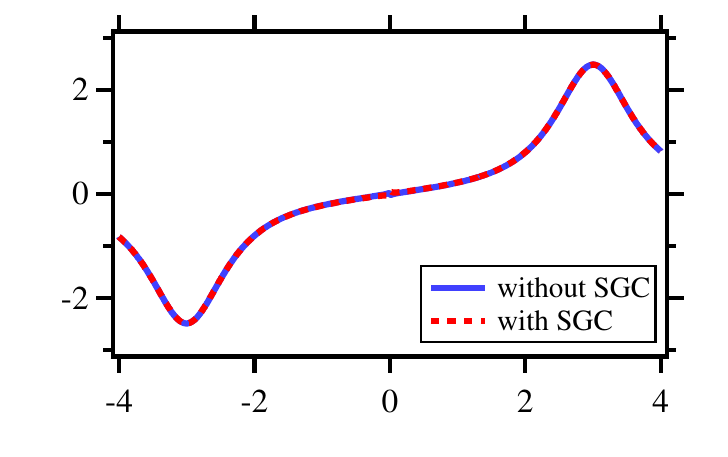}
	  \begin{picture}(0,0)
		\put(-200,40){\rotatebox{90}{$F/(\hbar k \Gamma) \times 10^{-1}$}}
		\put(-104,0){$kv/\Gamma$}
		\put(-223,125){(b)}
		\end{picture}

		\begin{tikzpicture}[overlay]
			\draw[lightgrey, thick] (0.3,2.85) ellipse [x radius=0.3, y radius=0.3];
		\end{tikzpicture}

		\includegraphics[width=0.8\linewidth]{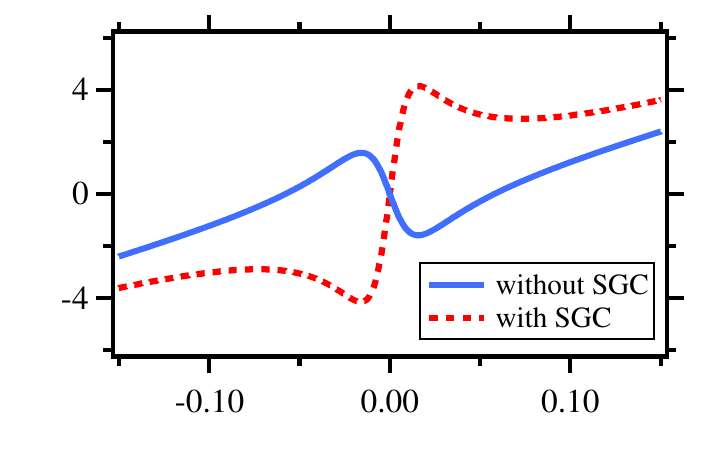}
	  \begin{picture}(0,0)
		\put(-200,40){\rotatebox{90}{$F/(\hbar k \Gamma) \times 10^{-3}$}}
		\put(-104,0){$kv/\Gamma$}
		\put(-223,125){(c)}
		\end{picture}
	  \caption{\label{forcevVel}(Color online) (a) Energy level diagram of the $F_g = 1 \rightarrow F_e=2$ atomic system. All the Zeeman energy levels are labelled using $\ket{i}$ notation. Clebsch–Gordan coefficeints are shown near the transitions. Velocity-dependent force curve for the $F_g = 1 \rightarrow F_e=2$ system is presented for (b) a large velocity range and (c) a small velocity range. The dashed red (solid blue) curve corresponds to the presence (absence) of SGC. The grey-circled portion in (b) is magnified and illustrated in (c). Parameters: $\Delta_{1 {\rightarrow} 2}=+3 \Gamma$ and $\Omega=\Gamma/\sqrt{2}$.}
\end{figure}

First, we analyze the cooling by two counter-propagating lasers with $\sigma^{+}$ and $\sigma^{-}$ polarizations for $F_g=1$ $\rightarrow$ $F_e=2$ with all the magnetic sub-levels (as shown in Fig. \ref{forcevVel} (a)) using density matrix calculations. The Hamiltonian, $H$ of the $8$ level system under the electric-dipole and rotating wave approximation and in the rotating frame is expressed as: $H=H_A+H_I$, where
\begin{align}
\label{eq1}
H_A= \,&2kv\ket{3}\bra{3}+(kv-\Delta)\ket{4}\bra{4}+(kv-\Delta)\ket{5}\bra{5}\nonumber\\
&-(kv+\Delta)\ket{6}\bra{6}-(kv+\Delta)\ket{7}\bra{7}\nonumber\\
&-(3kv+\Delta)\ket{8}\bra{8}
\end{align}
and
\begin{align}
\label{eq2}
H_I= \,&\frac{1}{2}\hbar\Omega\big\{\ket{1}\bra{4}+\frac{1}{\sqrt{6}}\ket{1}\bra{6}+\frac{1}{\sqrt{2}}\ket{2}\bra{5}+\frac{1}{\sqrt{2}}\ket{2}\bra{7}\nonumber\\
&+\frac{1}{\sqrt{6}}\ket{3}\bra{5}+\ket{3}\bra{8}\big\}+h.c.
\end{align}
Here, $\Delta$, $\Omega$ and $k$ denote the detuning, rabi frequency and the magnitude of the wave-vector of the lasers respectively, $v$ is the velocity of the atoms. The atom-field interaction is described by writing the Liouville-von Neumann equation for the density matrix,
\begin{align}
\label{Linb}
&\dot{\rho}=-\frac{i}{\hbar}[H, \rho]+ L(\rho)
\end{align}
Here, $L(\rho)$ accounts for the spontaneous decay of atoms via various channels. 64 simultaneous differential equations are obtained. Only equations for $\dot{\rho}_{12}$, $\dot{\rho}_{13}$ and $\dot{\rho}_{23}$ have the effect of SGC, which are given below. The terms appearing due to SGC are underlined.

\begin{align}
\dot{\rho}_{12}=\,&-\frac{i}{2}\Omega(\rho _{42}+\frac{1}{\sqrt{6}}\rho _{62})+\frac{i}{2\sqrt{2}}\Omega^{*}(\rho _{15}+\rho _{17})\nonumber\\
&\underline{+\Gamma(\frac{1}{\sqrt{2}}\rho _{45}+\frac{1}{\sqrt{3}}\rho _{56}+\frac{1}{2\sqrt{3}}\rho _{67})}\\ 
\dot{\rho}_{13}=\,&-\frac{i}{2}\Omega(\rho _{43}+\frac{1}{\sqrt{6}}\rho _{63})+\frac{i}{2}\Omega^{*}(\frac{1}{\sqrt{6}}\rho _{16}+\rho _{18})\nonumber\\
&+2ikv\rho _{13}\underline{+\Gamma(\frac{1}{\sqrt{6}}\rho _{46}+\frac{1}{2}\rho _{57}+\frac{1}{\sqrt{6}}\rho _{68})}\\ 
\dot{\rho}_{23}=\,&-\frac{i}{2\sqrt{2}}\Omega(\rho _{53}+\rho _{73})+\frac{i}{2}\Omega^{*}(\frac{1}{\sqrt{6}}\rho _{26}+\rho _{28})\nonumber\\
&+ikv\rho _{23}\underline{+\Gamma(\frac{1}{2\sqrt{3}}\rho _{56}+\frac{1}{\sqrt{3}}\rho _{67}+\frac{1}{\sqrt{2}}\rho _{78})}
\end{align}
The remaining equations are presented in \ref{F12}. They are numerically solved and the force experienced by the atom is evaluated from the absorption of the $\sigma^{+}$ and $\sigma^{-}$ light, following a similar approach as in \cite{Dalibard1989JosaB,Chang2002PhyRep,rajnandan2024continuous}. The force on atom can be given by following expression,

\begin{align}
	\label{Force}
	F_{\mathrm{damp}}= \,&\hbar k \Omega \mathrm{~ Im}\big[(\rho_{14}-\rho_{38})+\frac{1}{\sqrt{2}}(\rho_{25}-\rho_{27})\nonumber\\
	&+\frac{1}{\sqrt{6}}(\rho_{36}-\rho_{16})
	\big]
\end{align} 

In Fig. \ref{forcevVel} (b), we present the force vs velocity plot for a large velocity range under the influence of blue detuned lasers. The red dashed (blue solid) curve corresponds to the force in the presence (absence) of SGC. Both the curves reveal indistinguishable Doppler force profiles for large velocity ranges. The analysis indicates that atoms with positive (negative) velocities encounter positive (negative) forces, leading to heating in the presence of blue detuned lasers. This force vs velocity curves undergo sign reversal for negative detuning, indicative of the well known Doppler cooling.

Zooming into the grey-circled region in Fig. \ref{forcevVel} (b), Fig. \ref{forcevVel} (c) provides a closer examination of the small velocity range. SGC exerts a substantial influence on the force vs velocity behavior, particularly for lower velocities. In the presence of SGC, atoms with positive (negative) velocities experience a more pronounced positive (negative) force compared to the Doppler force alone, contributing to enhanced heating. Conversely, diminishing the impact of SGC results in a sign reversal in the slope of the force vs velocity curve, indicating the onset of blue-detuned cooling in a type-I system. These curves for force vs velocity flip signs for negative detuning, leading to polarization gradient cooling (heating) in the presence (absence) of SGC.

\section{Experimental setup}

	The experimental set-up comprises of one commercially available (Toptica) 420 nm (blue) external cavity diode laser (ECDL) and two home-assembled 780 nm (IR) ECDLs. Polarization spectroscopy is employed for the IR MOT laser's frequency stabilization \cite{rajnandan2023}, while saturation absorption spectroscopy is used for the IR repumper laser and the blue laser (similar to our previous experiments \cite{rajnandan2024continuous,rajnandan2024direct}). Four sets of beams $L_1$, $L_2$, $L_3$ and $L_4$ are derived which are IR MOT (5S$_{1/2}$, F $=2$ $\rightarrow$ 5P$_{3/2}$, F $=3$), IR repumper (5S$_{1/2}$, F $=1$ $\rightarrow$ 5P$_{3/2}$, F $=$ 1 or 2), red detuned blue MOT (5S$_{1/2}$, F $=2$ $\rightarrow$ 6P$_{3/2}$, F $=3$) and blue detuned blue MOT (at D$_1$ or D$_2$ line depending on the configuration) beams respectively. All these beams are switched on/off using AOMs. Each beam is further divided into 3 beams, overlapped, co-propagated, made circularly polarized using dual $\lambda/4$ wave plates, expanded to diameter of 25 mm, sent to the MOT chamber and retro-reflected back using dual $\lambda/4$ wave plate and mirror. In each arm, polarization of $L_1$ and $L_3$ are same and $L_2$ and $L_4$ are same but orthogonal to $L_1$ and $L_3$ .
	
	Atomic vapor is introduced into the MOT chamber by passing 2.15 A current to the dispenser (AlfaSource AS-Rb-0090-2C). First, the IR MOT is loaded for 2 s at quadruple magnetic field ($B'$) of 12.5 G/cm by switching on $L_1$ (power, 50 mW and detuning $-10$ MHz from 5S$_{1/2}$, F $=2 \rightarrow$ 5P$_{3/2}$, F $=3$ transition) and $L_2$ (power, $P_{1\to2}=33$ mW and detuning, $\Delta_{1\to2}/2\pi=+40$ MHz from 5S$_{1/2}$, F $=1 \rightarrow$ 5P$_{3/2}$, F $=2)$ transition. Power of the $L_1$ is lowered by 5 times and is waited for 4 ms to lower the temperature. Number of atoms, $N$ in the IR MOT is $\sim1.3\times10^{8}$ and the temperature, $T$ is $\sim 2$ mK. Then it is transferred to the red detuned blue MOT by switching off the $L_1$ beam and switching on the $L_3$ beam (power 26 mW and detuning $-7$ MHz from 5S$_{1/2}$, F $=2 \rightarrow$ 6P$_{3/2}$, F $=3$ transition). After 4 ms, power of $L_3$ beam is reduced to 10 mW and detuning is ramped to $\Delta_{2\to3}/2\pi=-3$ MHz in 5 ms. After 20 ms of hold time, $L_2$, $L_3$ and $B'$ are switched off. $N$ and $T$ are measured from the time of flight (TOF) method using absorption imaging at 5S$_{1/2}$, F $=2 \rightarrow$ 5P$_{3/2}$, F $=3$ transition on CMOS camera with exposure time of 100 $\mu$s. 
	
\begin{figure*}[t]
	\centering
	\hfill
	\includegraphics[height=4.5 cm]{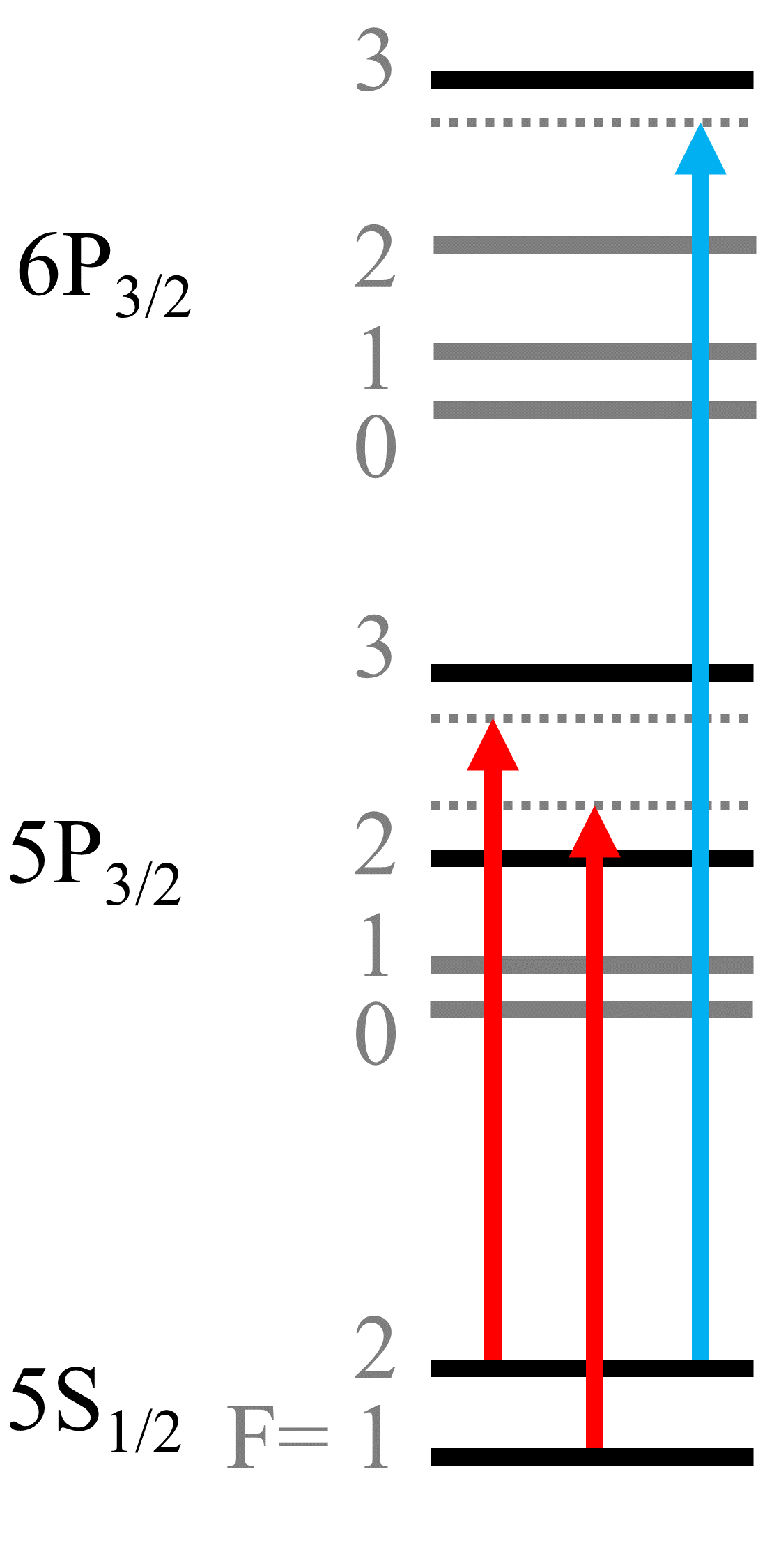}
	\begin{picture}(0,0)
		\put(-65,122){(a)}
	\end{picture}
	\hfill
	\includegraphics[width=0.26\linewidth]{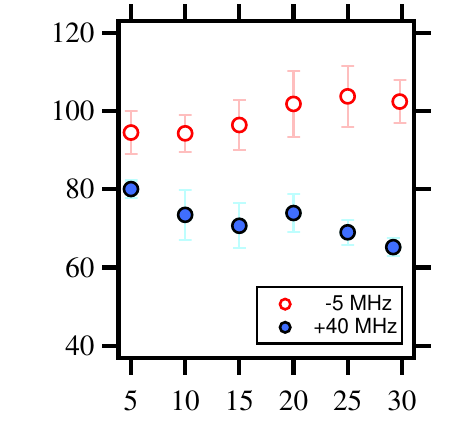}
	\begin{picture}(0,0)
		\put(-142,122){(b)}
					\put(-140,55){\rotatebox{90}{$T ~(\mu \textrm{K})$}}
					\put(-85,-3){$P_{1\to2}$ (mW)}
	\end{picture}
	\hfill
		 \includegraphics[width=0.26\linewidth]{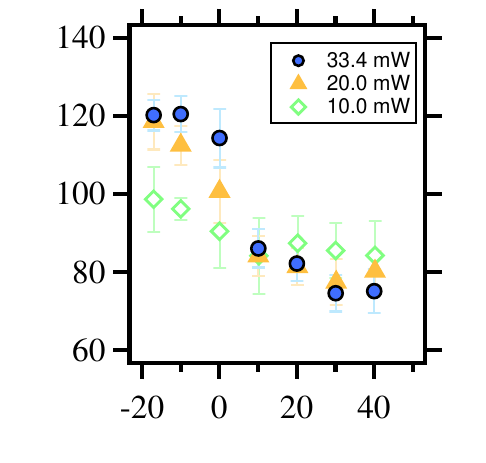}	
		 \begin{picture}(0,0)
			\put(-142,122){(c)}
				\put(-140,55){\rotatebox{90}{$T ~(\mu \textrm{K})$}}
				\put(-90,-3){$\Delta_{1\to2}/2\pi$ (MHz)}
		\end{picture}
		\hfill
			\caption{\label{rep12blue23}(Color online) (a) Relevant energy levels for studying the effect of repumper laser addressing $F=1\to2$ transition. (b) Temperature vs power of the repumper laser when $\Delta_{1\to2}/2\pi=+40$ MHz (filled blue circle) and $-5$ MHz (unfilled red circle). (c) Temperature vs detuning of the repumper laser when P$_{1\to2}=10$ mW (unfilled green diamond), $20$ mW (filled yellow triangle) and $33.4$ mW (filled blue circle).}
	 \end{figure*}

	 \begin{figure*}[t!]
		\centering
		\hfill
		\includegraphics[height=4.5 cm]{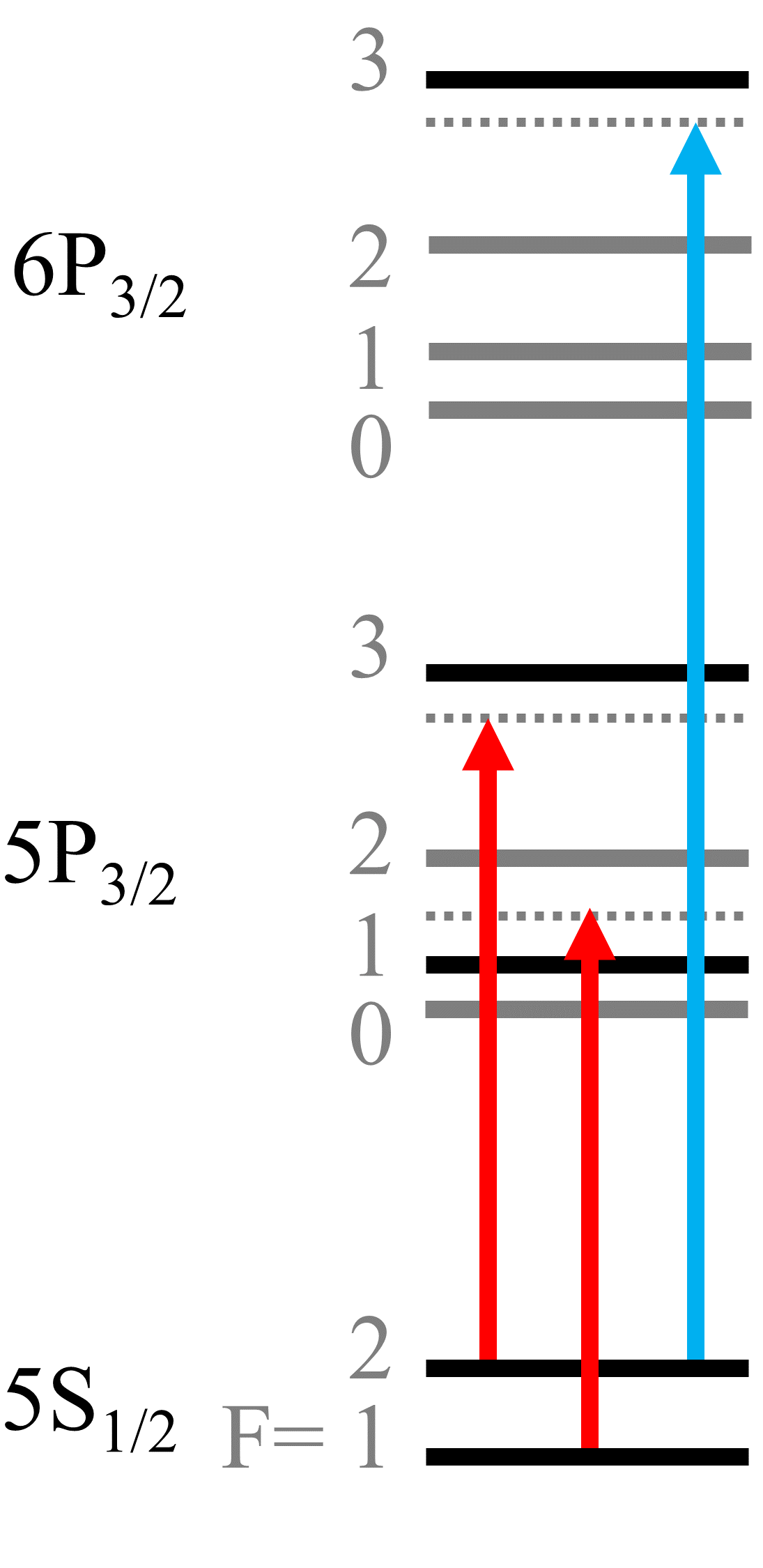}
		\begin{picture}(0,0)
			\put(-65,122){(a)}
		\end{picture}
		\hfill
				\includegraphics[width=0.26\linewidth]{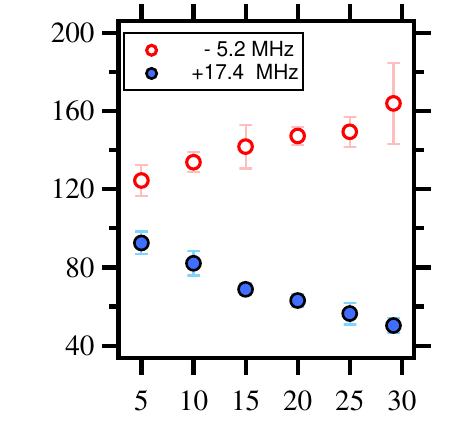}
				\begin{picture}(0,0)
					\put(-142,122){(b)}
					\put(-140,55){\rotatebox{90}{$T ~(\mu \textrm{K})$}}
					\put(-85,-3){$P_{1\to1}$ (mW)}
				\end{picture}
		\hfill
			 \includegraphics[width=0.26\linewidth]{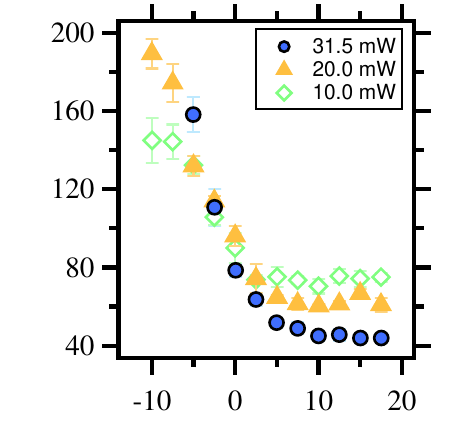}
			 \begin{picture}(0,0)
				\put(-142,122){(c)}
				\put(-140,55){\rotatebox{90}{$T ~(\mu \textrm{K})$}}
				\put(-90,-3){$\Delta_{1\to1}/2\pi$ (MHz)}
			\end{picture}	
			\hfill
		\caption{\label{rep11blue23}(Color online) (a) Relevant energy levels for studying the effect of repumper laser addressing $F=1\to1$ transition. (b) Temperature vs power of the repumper laser when $\Delta_{1\to1}/2\pi=+17.4$ MHz (filled blue circle) and $-5.2$ MHz (unfilled red circle). (c) Temperature vs detuning of the repumper laser when P$_{1\to1}=10$ mW (unfilled green diamond), $20$ mW (filled yellow triangle) and $31.5$ mW (filled blue circle). }
		 \end{figure*}

\section{Results and discussion}	

We first study the effect of the orthogonally polarized IR repumper laser ($F_g=1$ $\rightarrow$ $F_e=$ 2 ) on the IR MOT by changing the power ($P_{1\to2}$) at $\Delta_{1\to2}/2\pi=+40$ MHz. We observe that with increase in $P_{1\to2}$ from 0.5 mW to 33 mW, $T$ of the IR MOT increases from 1 mK to 2 mK. $N$ also increases and saturates to $1.3\times 10^{8}$. We then vary $\Delta_{1\to2}/2\pi$ from $-20$ MHz to $+40$ MHz at $P_{1\to2} = 33$ mW. We observe no significant change in $N$ and $T$. This is because the IR MOT laser is driving a close transition and only small fraction of atoms are lost due to off-resonant excitation.	 

Now, we study the effect of IR repumper on the red detuned blue MOT. As the blue transition is open with poor branching ratio, atoms decay continuously to the lower ground state, F$_g$=1.  
Fig. \ref{rep12blue23} (b) and (c) show the variation of the $T$ with $P_{1\to2}$ and $\Delta_{1\to2}/2\pi$ respectively for the configuration shown in \ref{rep12blue23} (a). Here, the $L_3$ is at $-3$ MHz detuned to the 5S$_{1/2}$, F$=2 \rightarrow$ 6P$_{3/2}$, F$=3$ transition. First, the repumper laser is kept at +40 MHz detuned to the  F$=1 \rightarrow$ F$=2$  transition (which is also a type-I transition) and the power of the repumper laser is varied. We observe that with increase in P$_{1\to2}$ from 5 mW to 30 mW, $T$ decreases from $80~\mu$K to $65~\mu$K as shown in Fig. \ref{rep12blue23} (b) by filled blue circle points. This is opposite to the case when the repumper laser is red detuned. When $\Delta_{1\to2}/2\pi=-5$ MHz, $T$ increases from $90~\mu$K to $105~\mu$K as shown by unfilled red circle points (a). In both the cases, $N$ increases with the increase in P$_{1\to2}$ and then saturates to $1.1\times 10^{8}$.

Next, we study the behavior of the red detuned blue MOT with the detuning of the repumper laser, $\Delta_{1\to2}/2\pi$ at three different power: 10 mW, 20 mW and 33 mW, as shown in Fig. \ref{rep12blue23} (c) by unfilled green diamond, filled yellow triangle and filled blue circle points respectively. When $\Delta_{1\to2}/2\pi$ is varied from $-20$ MHz to $+40$ MHz at P$_{1\to2}=33$ mW, $T$ decreases from $120~\mu$K to $75~\mu$K and then saturates. Similar pattern is observed at P$_{1\to2}=10$ mW and 20 mW. We observe that blue detuned laser cooling works even at type-I transition and the blue detuned repumper laser help the narrow line MOT at blue transition to reach lower temperature than the red detuned repumper laser.
				
Same study is done for the configuration shown in Fig. \ref{rep11blue23} (a), where the repumper laser is $-5.2$ MHz red detuned to the  F$=1 \rightarrow$ F$=1$  transition, which is a type-II transition. Similar to the previous case, $T$ increases from $120~\mu$K to $150~\mu$K when the power of the repumper laser (P$_{1\to1}$) is increased from 5 mW to 30 mW (as shown in red in Fig. \ref{rep11blue23} (b)). In the blue detuned repumper laser configuration i.e. when $\Delta_{1\to1}/2\pi=+17.4$ MHz, $T$ shows a decreasing trend from $90~\mu$K to $44~\mu$K with increase in P$_{1\to1}$ from 5 mW to 30 mW. In both the cases, $N$ increases with P$_{1\to1}$ and then saturates to $1.1\times 10^{8}$.

We study the effect of repumper laser detuning ($\Delta_{1\to1}$) on the blue MOT at three different powers: 10 mW, 20 mW and 31.5 mW (as shown in Fig. \ref{rep11blue23} (b) by unifilled green diamond, filled yellow triangle and filled blue circle points respectively). At P$_{1\to1}=31.5$ mW, $T$ decreases from $> 200~\mu$K to $44~\mu$ K and reaches saturation as the $\Delta_{1\to1}/2\pi$ is changed from $-10$ MHz to $+17$ MHz. Similar trend is observed when P$_{1\to1}=10$ mW and $20$ mW, but $T$ saturates to higher value.
	 
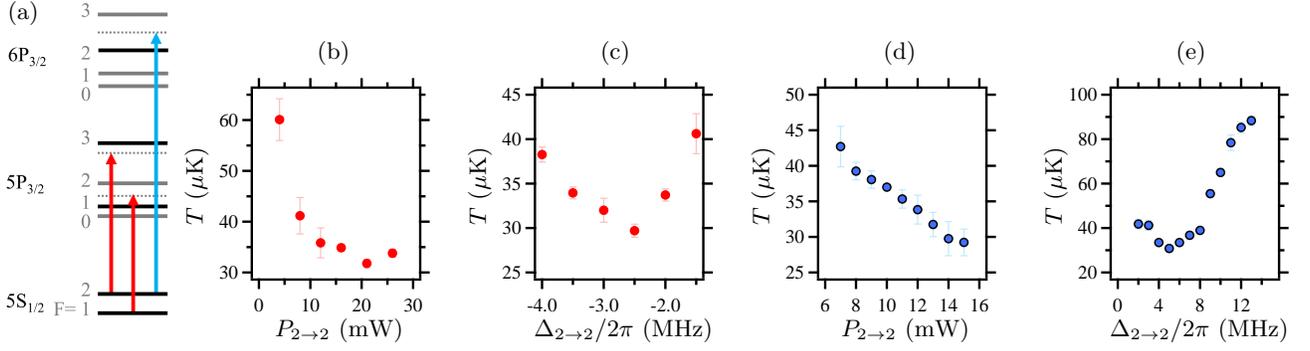
\begin{figure*}
	\centering
	\includegraphics[height=4.5 cm]{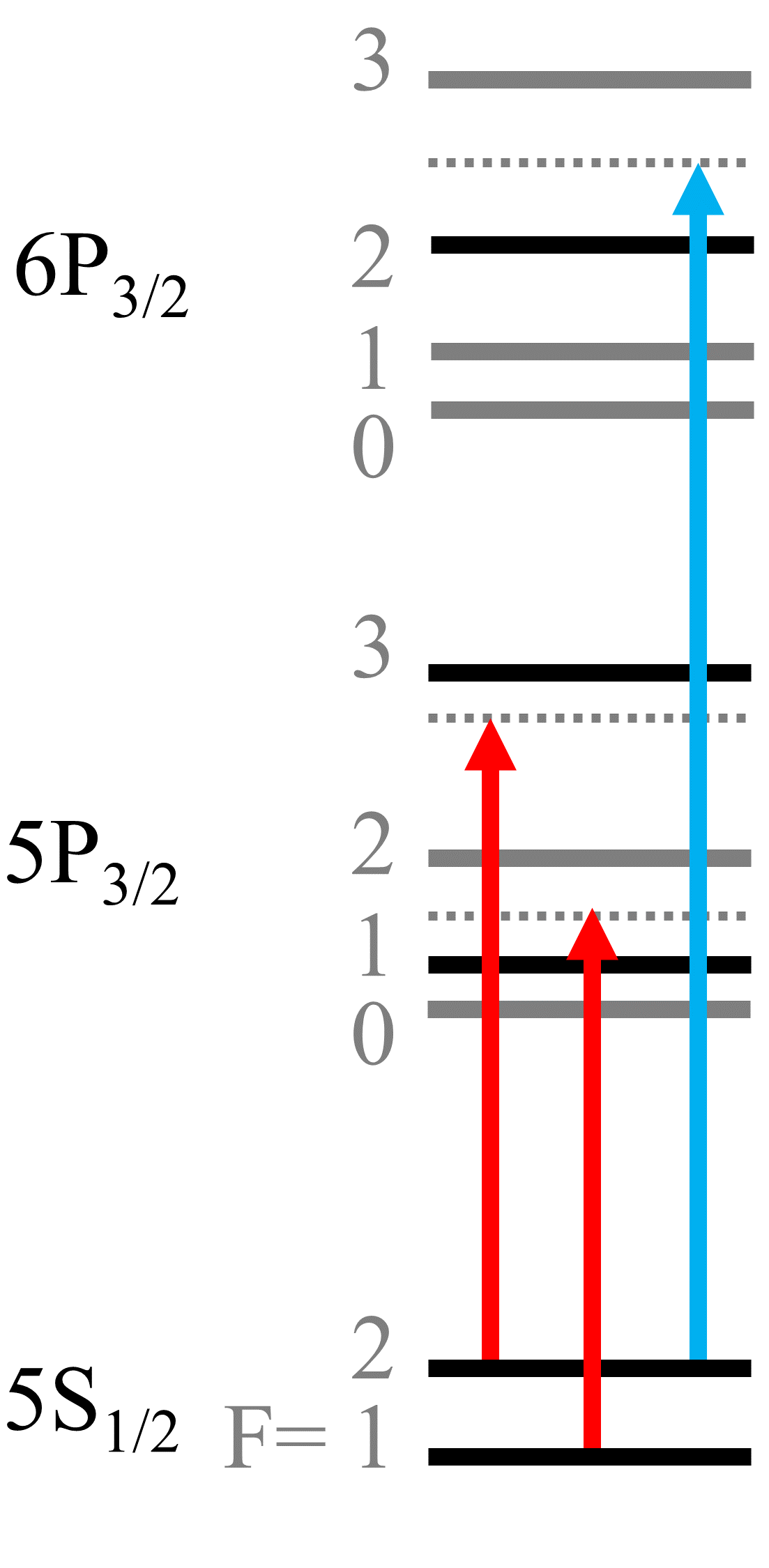}
		\begin{picture}(0,0)
			\put(-65,120){(a)}
		\end{picture}
	\includegraphics[height=3.5 cm]{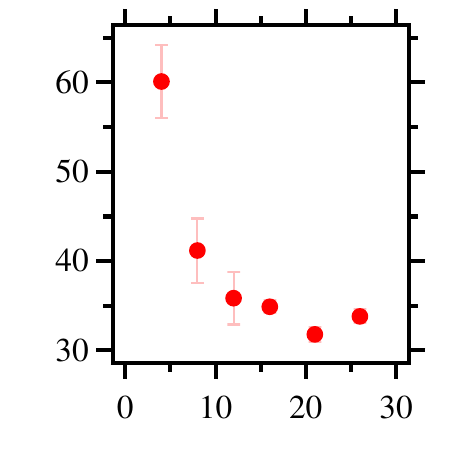}
	\begin{picture}(0,0)
		\put(-55,105){(b)}
		\put(-105,42){\rotatebox{90}{$T ~(\mu \textrm{K})$}}
		\put(-71,0){$P_{2\to2}$ (mW)}
	\end{picture}
	\includegraphics[height=3.5 cm]{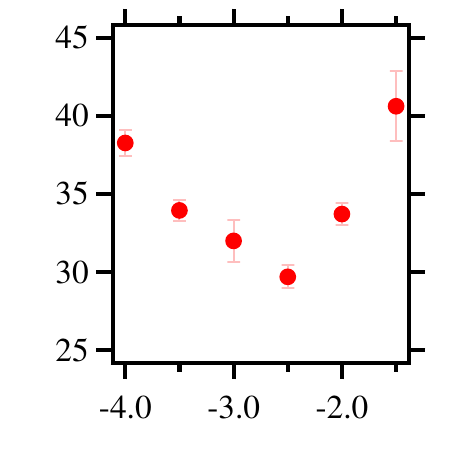}
	\begin{picture}(0,0)
		\put(-55,105){(c)}
		\put(-105,42){\rotatebox{90}{$T ~(\mu \textrm{K})$}}
		\put(-80,0){$\Delta_{2\to2}/2\pi$ (MHz)}
	\end{picture}
	\includegraphics[height=3.5 cm]{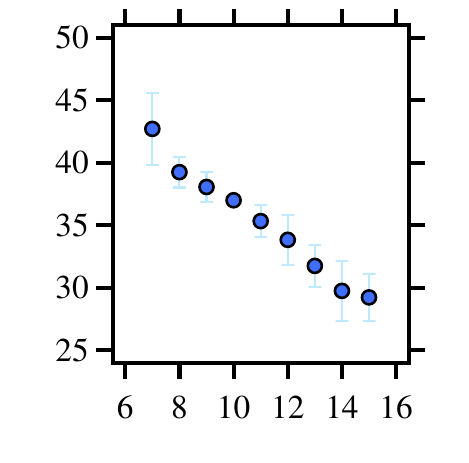}
	\begin{picture}(0,0)
		\put(-55,105){(d)}
		\put(-105,42){\rotatebox{90}{$T ~(\mu \textrm{K})$}}
		\put(-71,0){$P_{2\to2}$ (mW)}
	\end{picture}
	\includegraphics[height=3.5 cm]{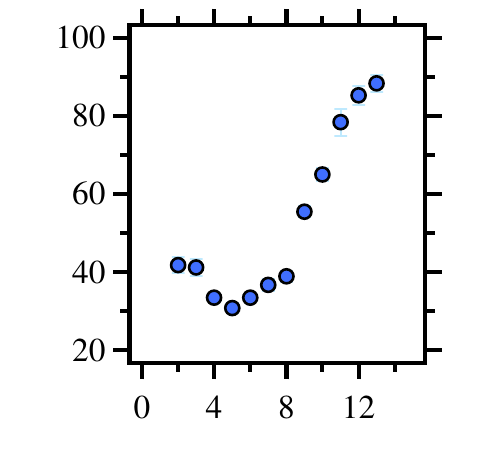}
	\begin{picture}(0,0)
		\put(-55,105){(e)}
		\put(-105,42){\rotatebox{90}{$T ~(\mu \textrm{K})$}}
		\put(-80,0){$\Delta_{2\to2}/2\pi$ (MHz)}
	\end{picture}

	\caption{\label{rep11blue22}(Color online) (a) Relevant energy levels for studying the effect of the blue laser addressing $F=2\to2$ transition. (b) Temperature vs power of the 420 nm laser ($L_3$) when $\Delta_{2\to2}/2\pi=-2$ MHz. (c) Temperature vs detuning of the 420 nm laser ($L_3$) when P$_{2\to2}=26$ mW. (d) Temperature vs power of the 420 nm blue detuned laser ($L_4$) when $\Delta_{2\to2}/2\pi=+5$ MHz. (e) Temperature vs detuning of the 420 nm blue detuned laser ($L_4$) when P$_{2\to2}=15$ mW. In (b)-(e), $\Delta_{1\to1}/2\pi=+14$ MHz and P$_{1\to1}=30$ mW.}
	 \end{figure*}

Next, we demonstrate the red detuned blue MOT using 5S$_{1/2}$, F$=2 \rightarrow$ 6P$_{3/2}$, F$=2$ transition (type-II MOT) with repumper laser at $30$ mW and $+14$ MHz blue detuned from the 5S$_{1/2}$, F$=1 \rightarrow$ 5P$_{3/2}$, F$=1$ transition. We study the effect of the power (P$_{2\to2}$) of the red detuned blue MOT beam at $\Delta_{2\to2}/2\pi=-2.5$ MHz and observe that with increase in P$_{2\to2}$ from 5 mW to 20 mW, $T$ decreases from $60~\mu$K to $32~\mu$K and then increases after 20 mW, as show in Fig. \ref{rep11blue22} (b). Then we study the effect of $\Delta_{2\to2}$ on the blue MOT at P$_{2\to2}=26$ mW and observe that $T$ initially decreases from $38~\mu$K to $32~\mu$K as the $\Delta_{2\to2}/2\pi$ is changed from $-4$ MHz to  $-2.5$ MHz and then increases to $>40~\mu$K as the detuning is changed towards zero (as shown in Fig. \ref{rep11blue22} (c)). As compared to the minimum temperature of the red detuned blue MOT at 5S$_{1/2}$, F$=2 \rightarrow$ 6P$_{3/2}$, F$=3$ transition (44 $\mu$K), $T$ of the red detuned blue MOT at 5S$_{1/2}$, F$=2 \rightarrow$ 6P$_{3/2}$, F$=2$ transition (32 $\mu$K) is lower.

We fix the parameters to load the red detuned blue MOT at 26 mW power and $-2.5$ MHz detuned from the 5S$_{1/2}$, F$=2 \rightarrow$ 6P$_{3/2}$, F$=2$ transition. We then demonstrate the blue detuned blue MOT using the same transition by transferring the atoms from the red detuned blue MOT. This is done by switching off the $L_3$ beams and switching on the $L_4$ beams. The quadruple magnetic field is also increased to 45 G/cm. We wait for 20 ms for the atoms to settle in the blue detuned blue MOT. First, we vary the power of the blue detuned blue beam (P$_{2\to2}$) from 7 mW to 15 mW at $\Delta_{2\to2}/2\pi=+4$ MHz and observe that $T$ decreases from $43~\mu$K to $31~\mu$K and saturates, as shown in Fig. \ref{rep11blue22} (c). We also reduced the blue detuned blue beam size by 4 times to confirm if we are limited by its power and observe no change in temperature at maximum available power.

\begin{table}
	\centering
	\caption{Temperature of the different MOTs. Detuning $(\Delta)$ of each laser from its corresponding transition is shown. IR repumper laser is driving the 5S$_{1/2}$, F$=1 \rightarrow$ 5P$_{3/2}$, F$=X$ transition, where $X=1$ or $2$. Blue MOT laser is driving the 5S$_{1/2}$, F$=2 \rightarrow$ 6P$_{3/2(1/2)}$, F$=X$ transition for the D$_{2(1)}$ MOT, where $X=1,2$ or $3$.}
	\label{Tcompiled}

	\begin{tabular}{ccccccc}
		\toprule
		\multicolumn{2}{c}{\textbf{IR Repumper}} & \multicolumn{2}{c}{\textbf{Blue MOT}} & \multicolumn{1}{c}{\textbf{T ($\mu$K)}} \\
		\cline{1-4}
		Transition & $\Delta$ (MHz) & Transition & $\Delta$ (MHz) &  \\
		\midrule
		$1\rightarrow 2$ & $+40$ & D$_2$, $2\rightarrow 3$ & $-3$ & $65\pm 2$ \\
		$1\rightarrow1$ & $+17$ & D$_2$, $2\rightarrow 3$ & $-3$ & $44\pm 2$ \\
		$1\rightarrow1$ & $+17$ & D$_2$, $2\rightarrow 3$ & $+2$ & $53\pm 2$ \\
		$1\rightarrow1$ & $+14$ & D$_2$, $2\rightarrow 2$ & $-2.5$ & $32\pm 1$ \\
		$1\rightarrow1$ & $+14$ & D$_2$, $2\rightarrow 2$ & $+5$ & $31\pm 1$ \\
		$1\rightarrow1$ & $+14$ & D$_1$, $2\rightarrow 2$ & $-2$ & $40\pm 1$ \\
		$1\rightarrow1$ & $+14$ & D$_1$, $2\rightarrow 2$ & $+5$ & $24\pm 1$ \\
		$1\rightarrow1$ & $+14$ & D$_1$, $2\rightarrow 1$ & $-2$ & $61\pm 2$ \\
		\bottomrule
	\end{tabular}
\end{table}

Next we study the effect of the detuning of the blue detuned blue beam ($\Delta_{2\to2}$) at fixed power, P$_{2\to2}=15$ mW. As the $\Delta_{2\to2}/2\pi$ is changed from $+13$ MHz to $+5$ MHz, $T$ decreases from $90~\mu$K to $31~\mu$K and then further increases to $>40~\mu$K as the frequency is changed towards resonance. We do not see any significant difference in temperature between the red detuned blue MOT and the blue detuned blue MOT with 5S$_{1/2}$, F$=2 \rightarrow$ 6P$_{3/2}$, F$=2$ transition.

We then carry out the same study for the blue MOT in the D$_1$ line at 5S$_{1/2}$, F$=2 \rightarrow$ 6P$_{1/2}$, F$=2$ transition, which is also a type-II open transition but of weaker transition strength as compared to the D$_2$ line. We observe that the minimum temperature of the red detuned blue MOT at D$_1$ line is $40~\mu$K. We then transfer the atoms to the blue detuned blue MOT at same line and observe that $T$ is lowered to $24~\mu$K. No. of atoms in the blue detuned blue MOT at D$_1$ transition also decreases to $5\times10^{7}$.

We also study other configurations of the blue MOT and minimum temperature achieved at different configurations are summarized in Table \ref{Tcompiled}. Unlike in \cite{Tarbutt2018PRA}, we have observed almost spherical shape of the atomic cloud with gaussian disctribution of atoms in all different configurations of the MOTs.

\section{Conclusion}
In summary, we observe that SGC plays a crucial role in laser cooling in the sub-Doppler regime. Prior to this work, it was known that there is no sub-Doppler cooling with blue-detuned lasers in type-I transition. In this work, we show that it is possible to achieve cooling with blue-detuned lasers but in the absence of SGC. As SGC is fragile to stray magnetic fields, we observe blue-detuned cooling even in type-I transition in MOT. For completeness, we also explore various combinations for the effectiveness of blue-detuned laser cooling in both type-I and type-II MOT configurations, achieving temperatures as low as $24~\mu$K in the D$_1$ MOT and $31~\mu$K in the D$_2$ MOT. 

\begin{acknowledgments}
The authors are grateful to Prof. T. N. Dey for fruitful discussions. RCD acknowledges the Ministry of Education, Government of India, for the Prime Minister's Research Fellowship (PMRF), and K.P. acknowledges funding from DST through Grant No. DST/ICPS/QuST/Theme-3/2019.
\end{acknowledgments}

\section*{Appendix}
\appendix
\setcounter{section}{1}
\subsection{$F=1\to2$ system}
\label{F12}
The remaining equations obtained using Density matrix analysis for the 8 level system are presented below.

$\\
\dot{\rho}_{11}=-\frac{i}{2}\Omega(\rho _{41}+\frac{1}{\sqrt{6}}\rho _{61})+c.c.+ \Gamma(\rho _{44}+\frac{1}{2}\rho _{55}+\frac{1}{6}\rho _{66})\\
\dot{\rho}_{22}=-\frac{i}{2\sqrt{2}}\Omega(\rho _{52}+\rho _{72})+c.c.+ \Gamma(\frac{1}{2}\rho _{55}+\frac{2}{3}\rho _{66}+\frac{1}{2}\rho _{77})\\
\dot{\rho}_{33}=-\frac{i}{2}\Omega(\frac{1}{\sqrt{6}}\rho _{63}+\rho _{83})+c.c.+\Gamma(\frac{1}{6}\rho _{66}+\frac{1}{2}\rho _{77}+\rho _{88})\\ 
\dot{\rho}_{44}=\frac{i}{2}\Omega\rho _{41}+c.c.-\Gamma\rho _{44}\\
\dot{\rho}_{55}=\frac{i}{2\sqrt{2}}\Omega\rho _{52}+c.c.-\Gamma\rho _{55}\\ 
\dot{\rho}_{66}=\frac{i}{2\sqrt{6}}\Omega(\rho _{61}+\rho _{63})+c.c.-\Gamma\rho _{66}\\ 
\dot{\rho}_{77}=\frac{i}{2\sqrt{2}}\Omega\rho _{72}+c.c.-\Gamma\rho _{77}\\
\dot{\rho}_{88}=\frac{i}{2\sqrt{2}}\Omega\rho _{83}+c.c.-\Gamma\rho _{88}\\ 
\dot{\rho}_{14}=\frac{i}{2}\Omega(\rho _{11}-\rho _{44}-\frac{1}{\sqrt{6}}\rho _{64})-i\,\left(\Delta-kv\right)\rho _{14}-\frac{1}{2}\Gamma\rho _{14}\\ 
\dot{\rho}_{15}=\frac{i}{2}\Omega(\frac{1}{\sqrt{2}}\rho _{12}-\rho _{45}-\frac{1}{\sqrt{6}}\rho _{65})-i\,\left(\Delta-kv\right)\rho _{15}-\frac{1}{2}\Gamma\rho _{15}\\ 
\dot{\rho}_{16}=\frac{i}{2}\Omega(\frac{1}{\sqrt{6}}\rho _{11}+\frac{1}{\sqrt{6}}\rho _{13}-\rho _{46}-\frac{1}{\sqrt{6}}\rho _{66})-i\,\left(\Delta+kv\right)\rho _{16}-\frac{1}{2}\Gamma\rho _{16}\\ 
\dot{\rho}_{17}=\frac{i}{2}\Omega(\frac{1}{\sqrt{2}}\rho _{12}-\rho _{47}-\frac{1}{\sqrt{6}}\rho _{67})-i\,\left(\Delta+kv\right)\rho _{17}-\frac{1}{2}\Gamma\rho _{17}\\ 
\dot{\rho}_{18}=\frac{i}{2}\Omega(\rho _{13}-\rho _{48}-\frac{1}{\sqrt{6}}\rho _{68})-i\,\left(\Delta+3kv\right)\rho _{18}-\frac{1}{2}\Gamma\rho _{18}\\ 
\dot{\rho}_{24}=\frac{i}{2}\Omega(\rho _{21}-\frac{1}{\sqrt{2}}\rho _{54}-\frac{1}{\sqrt{2}}\rho _{74})-i\,\left(\Delta-kv\right)\rho _{24}-\frac{1}{2}\Gamma\rho _{24}\\ 
\dot{\rho}_{25}=\frac{i}{2\sqrt{2}}\Omega(\rho _{22}-\rho _{55}-\rho _{75})-i\left(\Delta-kv\right)\rho _{25}-\frac{1}{2}\Gamma\rho _{25}\\ 
\dot{\rho}_{26}=\frac{i}{2}\Omega(\frac{1}{\sqrt{6}}\rho _{21}+\frac{1}{\sqrt{6}}\rho _{23}-\frac{1}{\sqrt{2}}\rho _{56}-\frac{1}{\sqrt{2}}\rho _{76})-i(\Delta+kv)\rho _{26}-\frac{1}{2}\Gamma\rho _{26}\\ 
\dot{\rho}_{27}=\frac{i}{2\sqrt{2}}\Omega(\rho _{22}-\rho _{57}-\rho _{77})-i(\Delta+kv)\rho _{27}-\frac{1}{2}\Gamma\rho _{27}\\ 
\dot{\rho}_{28}=\frac{i}{2}\Omega(\rho _{23}-\frac{1}{\sqrt{2}}\rho _{58}-\frac{1}{\sqrt{2}}\rho _{78})-i(\Delta+3kv)\rho _{28}-\frac{1}{2}\Gamma\rho _{28}\\ 
\dot{\rho}_{34}=\frac{i}{2}\Omega(\rho _{31}-\frac{1}{\sqrt{6}}\rho _{64}-\rho _{84})-i(\Delta+kv)\rho _{34}-\frac{1}{2}\Gamma\rho _{34}\\ 
\dot{\rho}_{35}=\frac{i}{2}\Omega(\frac{1}{\sqrt{2}}\rho _{32}-\frac{1}{\sqrt{6}}\rho _{65}+\rho _{85})-i(\Delta+kv)\rho _{35}-\frac{1}{2}\Gamma\rho _{35}\\ 
\dot{\rho}_{36}=\frac{i}{2}\Omega(\frac{1}{\sqrt{6}}\rho _{31}+\frac{1}{\sqrt{2}}\rho _{33}-\frac{1}{\sqrt{6}}\rho _{66}-\rho _{86})-i(\Delta+3kv)\rho _{36}-\frac{1}{2}\Gamma\rho _{36}\\ 
\dot{\rho}_{37}=\frac{i}{2}\Omega(\frac{1}{\sqrt{2}}\rho _{32}-\frac{1}{\sqrt{6}}\rho _{67}-\rho _{87})-i(\Delta+3kv)\rho _{37}-\frac{1}{2}\Gamma\rho _{37}\\ 
\dot{\rho}_{38}=\frac{i}{2}\Omega(\rho _{33}-\frac{1}{\sqrt{6}}\rho _{68}-\rho _{88})-i(\Delta+5kv)\rho _{38}-\frac{1}{2}\Gamma\rho _{38}\\ 
\dot{\rho}_{45}=\frac{i}{2\sqrt{2}}\Omega\rho _{42}-\frac{i}{2}\Omega^{*}\rho _{15}-\Gamma\rho _{45}\\
\dot{\rho}_{46}=\frac{i}{2\sqrt{6}}\Omega(\rho _{41}+\rho _{43})-\frac{i}{2}\Omega^{*}\rho _{16}-2ikv\rho _{46}-\Gamma\rho _{46}\\ 
\dot{\rho}_{47}=\frac{i}{2\sqrt{2}}\Omega\rho _{42}-\frac{i}{2}\Omega^{*}\rho _{17}-2ikv\rho _{47}-\Gamma\rho _{47}\\
\dot{\rho}_{48}=\frac{i}{2}\Omega\rho _{43}-\frac{i}{2}\Omega^{*}\rho _{18}-4ikv\rho _{48}-\Gamma\rho _{48}\\ 
\dot{\rho}_{56}=\frac{i}{2\sqrt{6}}\Omega(\rho _{51}+\rho _{53})-\frac{i}{2\sqrt{2}}\Omega^{*}\rho _{26}-2ikv\rho _{56}-\Gamma\rho _{56}\\ 
\dot{\rho}_{57}=\frac{i}{2\sqrt{2}}\Omega(\rho _{52}-\rho _{27})-2ikv\rho _{57}-\Gamma\rho _{57}\\
\dot{\rho}_{58}=\frac{i}{2}\Omega(\rho _{53}-\frac{1}{\sqrt{2}}\rho _{28})-4ikv\rho _{58}-\Gamma\rho _{58}\\ 
\dot{\rho}_{67}=\frac{i}{2\sqrt{2}}\Omega\rho _{62}-\frac{i}{2\sqrt{6}}\Omega^{*}(\rho _{37}+\rho _{17})-\Gamma\rho _{67}\\
\dot{\rho}_{68}=\frac{i}{2}\Omega\rho _{63}-\frac{i}{2\sqrt{6}}\Omega^{*}(\rho _{38}+\rho _{18})-2ikv\rho _{68}-\Gamma\rho _{68}\\ 
\dot{\rho}_{78}=\frac{i}{2}\Omega\rho _{73}-\frac{i}{2\sqrt{2}}\Omega^{*}\rho _{28}-2ikv\rho _{78}-\Gamma\rho _{78}
$

	
	\bibliography{BlueMOT}
	
	\end{document}